# HeartSiam: A Domain Invariant Model for Heart Sound Classification

Reza Yousefi Mashhoor
Electrical Engineering Dept.
Sharif University of Technology
Tehran, Iran
reza_yousefimashhoor@ee.sharif.edu

Ahmad Ayatollahi
Electrical Engineering Dept.
Iran University of Science and Technology
Tehran, Iran
Ayatollahi@iust.ac.ir

*Abstract —* Cardiovascular disease is one of the leading causes of death according to WHO. Phonocardiography (PCG) is a cost-effective, non-invasive method suitable for heart monitoring. The main aim of this work is to classify heart sounds into normal/abnormal categories. Heart sounds are recorded using different stethoscopes, thus varying in the domain. Based on recent studies, this variability can affect heart sound classification. This work presents a Siamese network architecture for learning the similarity between normal vs. normal or abnormal vs. abnormal signals and the difference between normal vs. abnormal signals. By applying this similarity and difference learning across all domains, the task of domain invariant heart sound classification can be well achieved. We have used the multi-domain 2016 Physionet/CinC challenge dataset for the evaluation method. *Results:* On the evaluation set provided by the challenge, we have achieved a sensitivity of 82.8%, specificity of 75.3%, and mean accuracy of 79.1%. While overcoming the multi-domain problem, the proposed method has surpassed the first-place method of the Physionet challenge in terms of specificity up to 10.9% and mean accuracy up to 5.6%. Also, compared with similar state-of-the-art domain invariant methods, our model converges faster and performs better in specificity (4.1%) and mean accuracy (1.5%) with an equal number of epochs learned.

## I. Introduction

Mechanical activity of the heart muscles produces heart sounds. They can be heard and recorded by digital stethoscopes. Fundamental heart sounds (FHSs) include the first heart sound (S1) and the second heart sound (S2), also known as lub dub. In addition to FHSs, other extra or unusual sounds existing in heart sounds are known as murmurs.

Due to the recent advancement in hardware and software technologies and the existence of electronic stethoscopes, automatic analysis of heart sound recordings has become more applicable. Despite the low accuracy and sensitivity of cardiac auscultation compared to other cardiac diagnostic methods, it is simple, cost-effective, and non-invasive.

The 2016 Physionet/CinC challenge [5] has provided many heart sound recordings ranging from 5 to 120 seconds. The challenge has provided train-set recordings in 6 folders, each from an existing database and recorded by a different stethoscope/sensor. Test-set of this challenge is not publicly available. However, a validation set consisting of a subset of the train-set is introduced. As shown in Fig. 1, most recordings belong to folder e and are domain unbalanced. As [2] proposes, this domain variability and domain imbalance of the data can affect the feature extraction step, causing a machine learning model to be biased toward the majority source of training data.

We have proposed a Siamese architecture in which a convolutional neural network (CNN) model is learned to output embeddings that make class similar heart sounds close to each other and class different heart sounds far from each other in Euclidean space irrelevant from their domain. After learning such a CNN, we utilize a K-nearest neighbor (KNN) model for the classification.

The following sections briefly discuss related work on heart sound classification. Section III explains the specific details of the Siamese architecture and the learning method. In section IV, we compare our method with the existing state-of-the-art method in different metrics. We conclude our work in section V.

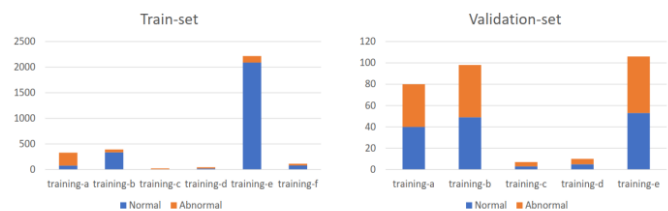

Fig. 1. Distribution of train-set and validation-set. Train-set is severely class and domain imbalanced. In contrast, validation-set is class balanced.

## II. Related Work

In the literature, heart sound classification usually consists of four steps: preprocessing, segmentation, feature extraction, and classification.[1]

### A. *Preprocessing:*

PCG signals are recorded in different environmental conditions, including in-house visits or hospitals. Several signal processing techniques are used for denoising. Digital filters are commonly used algorithm in denoising heart sound signals [6], [9], [10], [11], [12]. Band-pass filters are mostly used to reduce high and low-frequency noise in the signal. In [13], the authors applied the Savitzky-Golay filter, which fits a polynomial, to some local sample points to smooth the signal and remove tiny noises. Authors in [14] have analyzed different wavelets for signal decomposition to reduce noises. In [6], an algorithm is proposed to remove spikes in the signal.

*B. Segmentation:*

In this step, PCG signals are segmented into four parts: the first heart sound (S1), systole, second (S2), and diastole. Although this step is widely used in the literature, the classification task is feasible without segmentation [18], [19] [20].

Hidden Markov Model (HMM) is a popular machine learning algorithm used for heart sound segmentation [15], [16], [17]. In [6], the authors predict the most likely sequence of heart sounds based on the events' duration and the signal's envelope amplitude using a Hidden semi-Markov Model (HSMM). Authors in [7] used logistic regression instead of Gaussian distribution in [6] as emission probability and also extended the Viterbi algorithm and achieved state-of-the-art results. A two-dimensional U-net CNN architecture is proposed in [8], inspired by successful U-net architecture used in image segmentation. They also combined this method with work done in [7], considering outputs of CNN architecture as emission probabilities. These modifications further improved PCG segmentations.

*C. Feature Extraction*

Often feature extraction is used to convert raw signals into a better representation feature to improve the classification. Time domain features [9], [21], and frequency domain features [9] are commonly used. Mel-frequency cepstral coefficient (MFCC) is a popular feature in speech processing and is also used for processing heart sounds [18], [9], [13], [23].

In [2], the authors proposed a set of learnable filter banks trained to extract features well-suited for the classification task.

*D. Classification*

Machine learning classifiers such as AdaBoost [9], SVM [24], [25], KNN [11] are frequently used in the literature. With the recent advancement in deep learning, neural networks have gained more attention as classifiers. With the advantages of computationally efficient and parameter sharing, CNN is a popular and powerful architecture widely used in computer vision tasks. These networks have been used in several studies [9], [20], [2], [12], [13], [22], for heart sound classification.

Apart from various methods of heart sound classification, one of the critical challenges in this task, as described in the previous section, is the domain variability of the signals. Authors in [2] proposed a domain balanced batch training (DBT) which is a batch of training data with a balanced amount of data in each domain and each class. Appling DBT on the 2016 Physionet/CinC challenge dataset, they improved all domains' performances with slight degrading in domain e performance.

In [3], [4], Siamese architecture is introduced, which is two parallel weight-sharing neural networks (mostly CNNs) that learn to output embeddings that are similar if under the same class and dissimilar if under a different class. We have adopted this architecture to overcome the domain variability problem of the 2016 Physionet/CinC challenge data. To our knowledge, this is the first attempt to use Siamese networks in heart sound classification.

III. METHODOLOGY

The dataset used in this work is the multi-domain 2016 Physionet/CinC challenge dataset. As previously discussed, this dataset is highly domain and class imbalanced. We use preprocessing steps proposed in [6] for denoising and spike removal of the signals. Then for feature extraction, we decompose each signal into four frequency bands (25-45, 45-80, 80-200, and 200-400Hz) and use the segmentation method proposed in [7] to segment heart sounds into four parts.

We propose a Siamese network with three identical weight-shared CNNs to learn the similarity between two identical class signals (anchor and positive) and the difference between two different class signals (anchor and negative). The structure of the CNN can be seen in Fig. 3. We use the Euclidean distance as a similarity metric. If two signals are close to each other in Euclidean space, they are similar. We use triplet loss (1) as a loss function.

$$\sum_i^N \left[ \left\| f(x_i^a) - f(x_i^p) \right\|_2^2 - \left\| f(x_i^a) - f(x_i^n) \right\|_2^2 + \alpha \right] \quad (1)$$

Here $x_i^a$, $x_i^p$ and, $x_i^n$ is one sample of anchor, positive and negative respectively. $\alpha$ is a margin to ensure distance between anchor and negative gets sufficiently larger than anchor and positive after training.

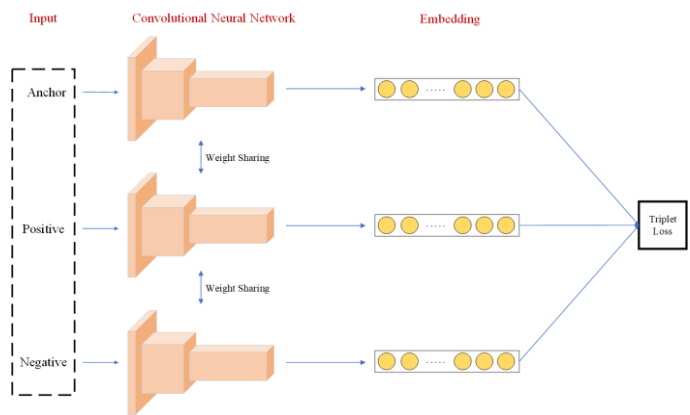

Fig. 2. Structure of the siamese network. Three CNNs are identical and their weights are shared so that modification in one network results the same in rest of them.

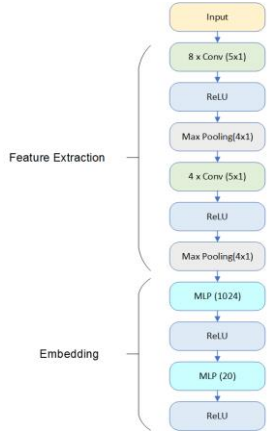

Fig. 3. Structure of the CNN model.

Fig. 4 shows the method of constructing a training dataset for Siamese architecture. Green and red squares represent preprocessed normal and abnormal data, respectively. To make a data block, we first choose one domain (E.g., domain A). Then we sample one normal and one abnormal data in this domain as an anchor (E.g., Green A1 and Red A1), and for each anchor, we sample one normal and one abnormal data in another domain as positive and negative randomly (E.g., B56 and B17). This results in a block of 12 triplets of data for one pair of normal-abnormal data in the selected domain (E.g., Block 1). N blocks result in training data of one domain (E.g., Training data A). N is a hyperparameter.

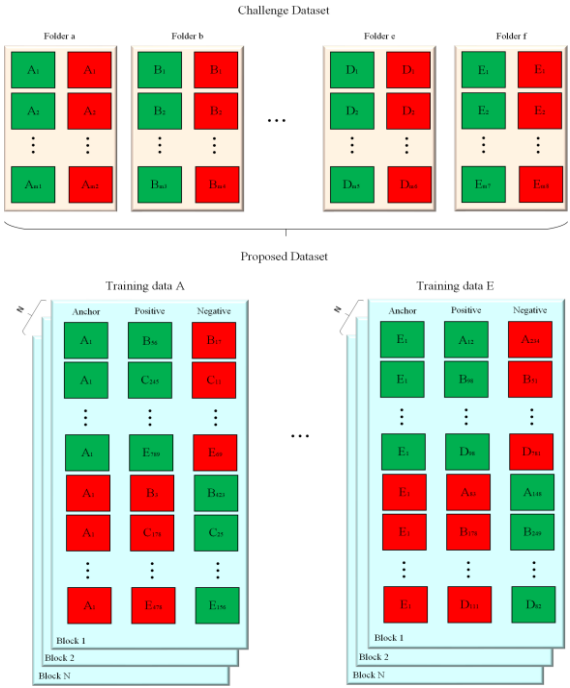

Fig. 4. Training dataset of Siamese architecture. Green and red squares represents normal and abnormal data. Challenge dataset is imbalanced, hence m1≠m2, m3≠m4 , … , m7≠m8.

After learning five Siamese models, we sample a class-balanced subset of all training data in the challenge dataset and feed it to one branch of each learned Siamese model to yield embeddings, and then for each set of embeddings, we learn a KNN model. For evaluation, similar to previous steps, we feed data into five parallel CNNs, and the KNN models predict data labels in each branch. The final label is the mean of all predicted labels.

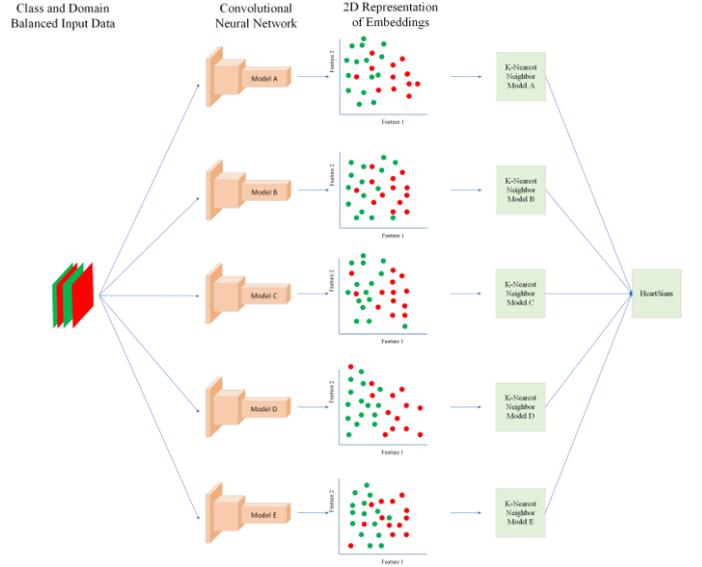

Fig. 5. Architecture of HeartSiam

## IV. EXPERIMENTS AND RESULTS

We use margin loss with α = 0.5 and Adam optimizer with learning rate = 10-4. We choose N=10000. We trained each CNN model for 30 epochs with batch size = 32. We have evaluated HeartSiam on the 2016 Physionet/CinC validation set and compared the results with methods proposed in [2]. For a signal, we have predicted each segment and then combined the predictions to form one prediction for each signal. Sensitivity, specificity, mean accuracy, and accuracy are the evaluation metrics. Table I compares the performance of HeartSiam with the previous state-of-the-art method [2]. The table's first row is the top-score method in the INTERSPEACH 2018 Computational Paralinguistics (ComParE) Challenge [27], which is used as one of the baseline models in [2]. It consists of bagged Support Vector Classification (SVC) ensembles trained on Bag of Audio Words (BoAW) representations. Further details on BoAW representation are available in [26]. The training data is balanced by upsampling the minority class. The second row of the table is the winning method in the 2016 Physionet/CinC challenge [9] which is another baseline model used in [2]. It has an FIR filter as input and employs a CNN to classify each segmented cardiac cycle. Based on the accuracies in each domain, these two methods are overfitted on the domain, e. The third and fourth row combines the DBT method described in section III after 30 and 200 epochs of learning, respectively. DBT method can modify models' performance in all other data domain accuracy with little degrading of domain e accuracy.

As can be seen in Table I, HeartSiam outperforms Potes-CNN and Gabor-BoAW-SVC (Upsamp.) in all metrics. Furthermore, with an equal number of epochs learned, HeartSiam surpasses the Potes-CNN DBT method. However, with a sufficient number of epochs, they have similar performance.

TABLE I. PERFORMANCE COMPARISON ON THE EVALUATION SET OF 2016 PYSIONET/CINC CHALLENGE

| Methods | Accuracy in data domains | | | | | | Sens. | Spec. | Macc |
|---|---|---|---|---|---|---|---|---|---|
| | *a* | *b* | *c* | *d* | *e* | *Avg* | | | |
| **Previous Methods** | | | | | | | | | |
| Gabor-BoAW-SVC (Upsamp.) | 50 | 57.14 | 57.14 | 60 | 99.01 | 64.66 | 72.18 | 67.12 | 69.65 |
| Potes-CNN | 60 | 63.26 | 71.43 | 40 | 100 | 66.94 | 82.61 | 64.38 | 73.5 |
| Potes-CNN DBT 30 epochs | 68.75 | 59.18 | 100 | 50 | 92.13 | 72.79 | 93.49 | 53.42 | 73.45 |
| Potes-CNN DBT | 71.25 | 70.41 | 85.71 | 66.66 | 97.5 | 78.31 | 87.68 | 71.91 | 79.79 |
| **Proposed Method** | | | | | | | | | |
| **HeartSiam** | 73.75 | 71.43 | 85.71 | 60 | 92.16 | 79.12 | 82.78 | 75.34 | 79.06 |

## V. CONCOLUSION

In this work, we have proposed a new model to classify heart sound recordings in the multi-domain 2016 Physionet/CinC challenge dataset into normal and abnormal. We suggest using a Siamese neural network with a CNN backend to produce embeddings that are close to each other if they belong to identical classes and are far apart otherwise. We proposed an algorithm to produce a semi-balanced training dataset. This method is used for heart sound classification but can be utilized in any other classification tasks with unbalanced data domains. Results show that our method is an easy learning algorithm and outperforms the previous state-of-the-art methods in sensitivity, specificity, and accuracy.

This work is implemented on a small amount of training data. However, we can improve model performance by collecting more binary labeled heart sound signals from various datasets and treating them as multi-domain input data. Heart sound segmentation is a time and resource-consuming step in this method. Based on the literature, end-to-end classification from unsegmented heart sound recordings are feasible; hence, it is suggested to be considered in future works.

## VI. REFFERENCES